\documentclass[12pt]{iopart}

\usepackage{graphicx}

\newcommand{\first}{1\textsuperscript{st} }
\newcommand{\second}{2\textsuperscript{nd} }
\newcommand{\third}{3\textsuperscript{rd} }

\begin{document}
\title{Precision measurements of quantum defects in the $n$P$_{3/2}$ Rydberg States of ${}^{85}$Rb}
\author{B.~Sanguinetti, H.~O.~Majeed, M.~L.~Jones and B.~T.~H.~Varcoe}
\address{School of Physics and Astronomy, University of Leeds, Leeds, LS2 9JT, UK}
\date{\today}
\maketitle
\begin{abstract}
Rydberg  States are used in our One Atom Maser experiment because they offer a large dipole moment and couple strongly to low numbers of microwave photons in a high Q cavity. Here we report the absolute frequencies of the P$_{3/2}$ states for principal quantum numbers $n=36$ to $n=63$. These measurements were made with a  three step laser excitation scheme. A wavemeter was calibrated against a frequency comb to provide accurate absolute frequency measurements over the entire range, reducing the measurement uncertainty to 1MHz. We compare the spectroscopic results with known frequency measurements as a test of measurement accuracy.
\end{abstract}

\section{Introduction}
Accurate knowledge of the excitation energies of Rydberg states is an important ingredient in increasing the accuracy of atomic structure calculations \cite{Drake91}. Experimentally, Rydberg atoms are also attractive because they offer a large dipole moment and a large number of closely spaced energy levels. 
Therefore there has been an increasing effort in the field of quantum information processing recently, where Rydberg states have been proposed for increasing the interaction strength between atoms \cite{Ryabtsev03,Walker08,Walker05,Jaksch00}, in cavity QED schemes for quantum computing \cite{Blythe06} and quantum optics research \cite{Walther06,Haroche06}. 

\begin{figure}[ht]
\begin{center}
\includegraphics[width=6cm]{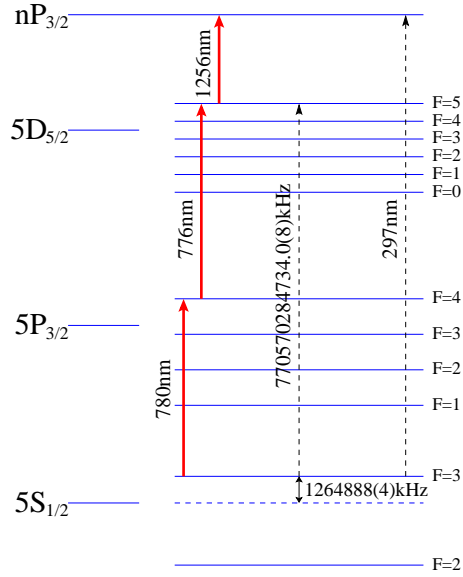}
\end{center}
\caption{Three level excitation scheme.\label{fig:levels}}
\end{figure}

For this experiment, the Rydberg states in ${{}^{85}}$Rb between $n=36$ and $n=63$ were excited using a three step laser excitation process as shown in Figure~\ref{fig:levels}. The steps are a 780.24nm transition 5S$_{1/2}$,~F=3 to 5P$_{3/2}$,~F=4, a 775.9nm laser resonant with the 5P$_{3/2}$,~F=4 to 5D$_{5/2}$,~F=5 transition and finally a 1256nm laser to excite $n$P Rydberg states. This method of excitation was preferred over a single UV step both because of the convenience and reliability of diode lasers and readily available optical fibres. The accuracy of the results was verified with a Menlo Systems frequency comb and we calculate values of the modified Rydberg-Ritz formula with significant improvements on previous results.

During the experiment the first two laser steps are stabilised to atomic reference transitions and the frequency of the transitions are measured with a frequency comb. 
The first step is locked to the 5S$_{1/2}$,~F=3 to 5P$_{3/2}$,~F=4  transition of ${}^{85}$Rb using a polarization spectrum \cite{Pearman2002}, which offers a dispersion curve without the need to modulate the laser frequency. 
Using the comb we determined that the locked laser linewidth is 300kHz with a 1 second Allan deviation $\leq$ 10kHz.

The second step laser is locked to the 5P$_{3/2}$,~F=4 to 5D$_{5/2}$,~F=5 transition. The locking signal is however derived from an enhanced transmission of the \first step laser light rather than directly from the \second step laser. Enhanced transparency of the \first step is used because it has a much higher signal to noise ratio at low intensities than any other signal and it therefore produces a better locking curve. The feedback signal is derived by applying a small modulation to the \second step and the locking curve is extracted using a lock-in amplifier. This is free from Doppler broadening because only zero Doppler class atoms are excited by the \first step laser, which is tightly locked to the Doppler free polarization spectrum. 

The third step laser is a broadly tunable `Stry' design \cite{Stry06} diode laser. This laser has a stability of better than 1MHz over a second, can be tuned accurately by a few GHz using a low voltage piezo actuator, and by nearly 100nm with a stepper motor. This laser is scanned over the 5D$_{5/2}$,~F=5 to $n$P$_{3/2}$ transition of interest by applying a voltage from a 16 bit optically isolated DAC to the piezo actuator.

\begin{figure}[ht]
\begin{center}
\includegraphics[width=6cm]{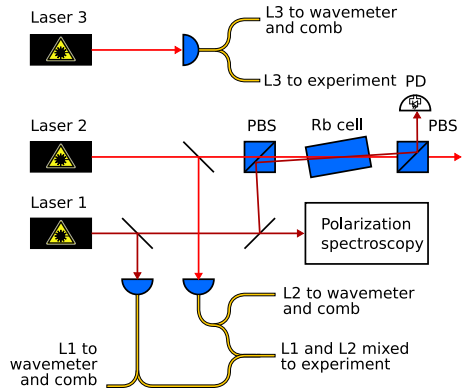}
\end{center}
\caption{A schematic showing the optical setup used in the 3 step excitation, including fibre optic delivery to both the diagnostic systems and the experiment itself.\label{fig:optics}}
\end{figure}

The beamline consists of a room temperature vacuum chamber ($P\simeq5\times 10^{-8}$mbar) in which we placed a rubidium dispenser \cite{Roach04}, a fibre coupled excitation region and a detector consisting of a set of field ionization plates with a small electric field ramp from 20 to 30V/cm and the resulting electrons are detected with a channel electron multiplier.
The lasers are launched into single mode optical fibre and passed into the vacuum using a fibre feedthrough. One single mode fibre (Nufern 780HP) carries the \first and \second step lasers and a second fibre (Corning SMF28) is used for the \third step laser (principally because the third step requires a larger core diameter fibre). The fibres are terminated in collimating lenses that produce a 2mm diameter beam. The output of the two fibres are arranged so that they are counter-propagating but meet the atomic beam in a perpendicular direction. A useful property of the counter-propagating lasers is the ability to check correct alignment as some light from the 780/776nm fibre is coupled into the 1260nm fibre when they are correctly aligned.

\begin{figure}[ht]
\begin{center}
\includegraphics[width=85.0mm]{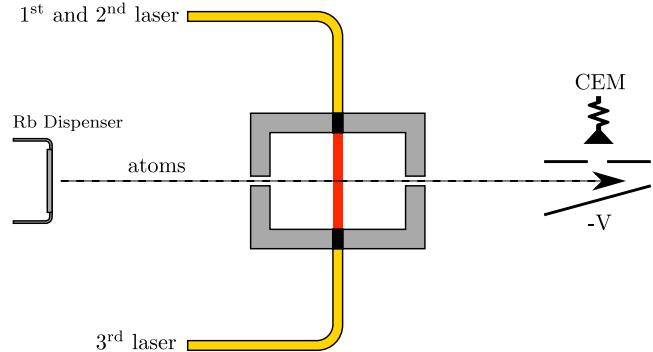} 
\end{center}
\caption{Beamline (not to scale): The entire beamline is housed in a vacuum chamber pumped with a turbo pump and an oil-free backing pump, reaching a background pressure of $5\times10^{-8}$mbar. The entire length is 50cm and includes a rubidium alkali metal dispenser, a shielded excitation region and field ionisation detector.\label{figure:beamline}}
\end{figure}

The self-referenced femtosecond frequency comb \cite{Udem02} was used to translate precision RF frequencies from an SRS FS725 Rubidium atomic clock to optical frequencies with an accuracy of $6\times 10^{-16}$ \cite{Kubina05} allowing direct frequency measurement with the accuracy of the RF reference to be made. The atomic clock provides an absolute frequency accuracy of better than $5\times 10^{-10}$ based on the manufacturers specification of the aging rates of the oscillator.
For a $10^{15}$Hz transition this corresponds to a potential absolute error of 500kHz.
The laser frequencies are measured using beat detection units that beat a closely lying comb line with the laser to be measured. The fixed frequency lasers can be simultaneously and continuously measured with this set up. To detect the frequency of the scanning third step laser we use a calibrated wavemeter. The wavemeter draws a comparison between two high finesse Fabry-Perot cavities and achieves an absolute accuracy of 60MHz. Using the frequency comb we have determined that the relative accuracy of the wavemeter, immediately following calibration, is $10^{-7}$ at 780nm (the comparison wavelength). 
During the experiment the wavemeter was continuously calibrated against the first step laser ensuring that the absolute frequency accuracy of the wavemeter was asserted with the frequency comb. Under this condition the wavemeter was disciplined to give an absolute accuracy of 1MHz. 
The frequency comb was also used to calibrate the accuracy of the target measurement conditions by observing the agreement between the comb and the wavemeter in the 1260nm range. 
At 1260nm the accuracy had degraded a little and the precision of 1260 measurements was 4MHz.
Finally spectral lines corresponding to the rubidium Rydberg transitions were acquired by scanning the \third step laser over a transition, and simultaneously recording the countrate and the \third step laser wavelength using a computer controlled data acquisition system. 
The experimental aim was to achieve an overall accuracy of 10MHz, as this the approximate linewidth of the transition and it is the maximum error level tolerated by the frequency comb electronics. 

The excitation region is a monolithic piece of aluminium machined to include both collimating holes for the atomic beam and fibre collimators, so that the alignment was as accurate as possible. The atomic beam is created with a 2cm long Rubidium dispenser oven \cite{Roach04}. The entire collimated cone is therefore filled by the beam and the finite angle of this cone will result in a finite Doppler broadening. This broadening is however much smaller than the linewidth of the \first step rubidium transition.
 
The stability of the \first laser is important for this experiment because it is used to calibrate the wavemeter, and as velocity selection for the second step lock. The Allan variance of this was measured to be below 10kHz for all the times relevant to the experiment. The second step lock uses an equivalent setup. The frequencies for these lasers were measured with the comb. In this paper we chose to use the combined two photon transition frequency from \cite{Nez93} which is known to 8kHz.

\begin{figure}[ht]
\begin{center}
\includegraphics[width=85.0mm]{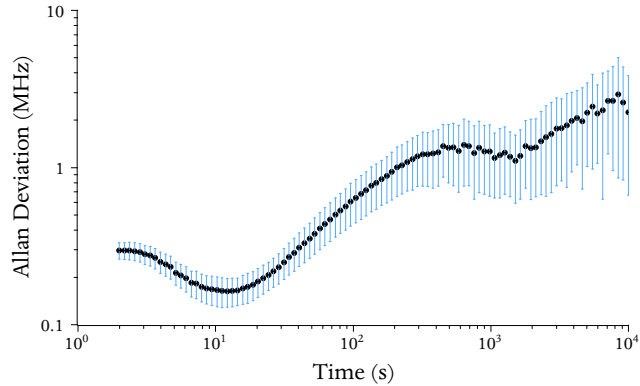}
\end{center}
\caption{Allan Deviation of the wavemeter. The data was taken over a period of twelve hours.\label{fig:wavemeter-allan-deviation}}
\end{figure}

The stability of the wavemeter was measured against the \first step laser while the laser was locked to the 5S-5P atomic transition (780nm). The absolute accuracy of the wavemeter remains below 2MHz for a few hours (see Figure \ref{fig:wavemeter-allan-deviation}) which is much longer than an individual scan but shorter than the entire experimental run, the wavemeter was therefore re-calibrated every hour. 
This calibration translates to 4MHz accuacy on the 1260nm laser (the last recorded digit in the figures). 
Systematic errors in the transfer of this calibration are therefore the largest error. 
The wavelength of the laser is read by the computer directly and therefore all plots and fitting routines use the absolute wavemeter reading as read at the time of the experiment.
Finally we determine the accuracy of the calibration by comparing the frequency of the combined transition from the 5S$_{1/2}\rightarrow$ 5D$_{5/2}$ states of 770~570~285(1)MHz which is consistent with the precision two photon measurment of Nez \cite{Nez93} who found 770~570~284~734(8)kHz. This demonstrates that the calibration procedure for the wavemeter was sufficient for the target accuracy.

\begin{table}[ht]
\begin{center}
\begin{tabular}{c|c|c|c|c}
	n  &\third step         & $E_n$               & $\delta$       & $\delta$ Error \\
	   & (MHz)       & (MHz)         &  & ($\times 10^{-5}$) \\ \hline
	36 &  236496706  &  1007068254  &  2.64187  &  2.3 \\
	37 &  236666310  &   1007237858  &  2.64179  &  2.5 \\
	38 &  236821728  &   1007393277  &  2.64170  &  2.7 \\
	39 &  236964479  &   1007536027  &  2.64175  &  2.9 \\
	40 &  237095926  &   1007667475  &  2.64177  &  3.2 \\
	41 &  237217235  &   1007788783  &  2.64173  &  3.4 \\
	42 &  237329406  &   1007900954  &  2.64176  &  3.7 \\
	43 &  237433360  &   1008004909  &  2.64162  &  4.0 \\
	44 &  237529853  &   1008101402  &  2.64160  &  4.3 \\
	45 &  237619595  &   1008191144  &  2.64156  &  4.6 \\
	46 &  237703191  &   1008274740  &  2.64163  &  5.0 \\
	47 &  237781211  &   1008352760  &  2.64151  &  5.3 \\
	48 &  237854117  &   1008425666  &  2.64154  &  5.7 \\
	49 &  237922362  &   1008493911  &  2.64148  &  6.1 \\
	50 &  237986322  &   1008557870  &  2.64155  &  6.5 \\
	51 &  238046352  &   1008617901  &  2.64167  &  6.9 \\
	52 &  238102791  &   1008674339  &  2.64144  &  7.3 \\
	53 &  238155879  &   1008727427  &  2.64161  &  7.8 \\
	54 &  238205906  &   1008777455  &  2.64159  &  8.2 \\
	55 &  238253103  &   1008824651  &  2.64139  &  8.7 \\
	56 &  238297662  &   1008869210  &  2.64139  &  9.2 \\
	57 &  238339780  &   1008911329  &  2.64148  &  9.8 \\
	58 &  238379637  &   1008951185  &  2.64158  &  10.3 \\
	59 &  238417400  &   1008988949  &  2.64141  &  10.9 \\
	60 &  238453197  &   1009024746  &  2.64151  &  11.5 \\
	61 &  238487172  &   1009058721  &  2.64151  &  12.1 \\
	62 &  238519445  &   1009090994  &  2.64151  &  12.7 \\
	63 &  238550123  &   1009121672  &  2.64165  &  13.4 \\

\end{tabular}
\end{center}
\caption{Measured frequencies for the nP$_{3/2}$ states and respective quantum defects. $E_n$ is measured from the centre of mass of the lower and upper states, and contains a small correction to the wavemeter calibration. The \third step data is reported exactly as measured.}
\label{table:results}
\end{table}

The detector field-ionizes the atoms and detects the subsequent electrons in a channel electron multiplier. The slew rate of the field in the detector is designed to be non-state selective, thus enabling the detector to collect electrons from any neighbouring Rydberg states may have been populated by black body radiation. The dark count rate of the detector was $\le$0.3Hz making the signal to noise ratio $S/N=300$.

\begin{figure}[ht]
\begin{center}
\includegraphics[width=85.0mm]{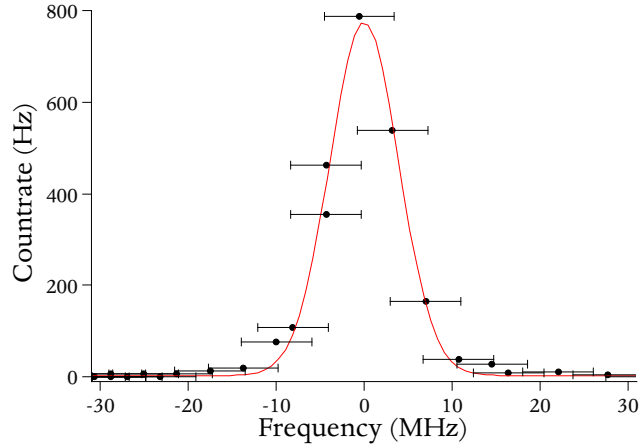}
\end{center}
\caption{Typical line, with a Lorentzian fit. Here the errors indicate the uncertainty in the absolute frequency as measured by the wavemeter.\label{figure:typical_line}}
\end{figure}

\begin{table}[ht]
\begin{center}
\begin{tabular}{|c|c|}
    \hline
	Wavemeter & 4MHz \\ \hline
	Angle & 4MHz \\ \hline
	\first step & 10kHz \\ \hline
	\second step & 500kHz \\ \hline
\end{tabular}
\caption{Error estimates}
\label{table:errors}
\end{center}
\end{table}
These systematic errors are not correlated resulting in a total systematic error of 5.7MHz. 
Subsequently we present the results of a scan from $n=36$ to $n=63$ in Table \ref{table:results}.

The measured frequencies, together with a calculated value of the quantum defect obtained from:
\begin{equation}
 E_n = E_i - \frac{R_\textrm{Rb} }{\left[ n-\delta(n)\right] ^2}
\end{equation}
are presented in table~\ref{table:results}. The values of $E_i$ were obtained by adding the measured \third step frequency to the 5S$_{1/2}\rightarrow$ 5D$_{5/2}$ transition frequency 770~570~285(1)MHz.
 We scanned over each line 10 times which revealed that the results presented below are repeatable to 1MHz. 
 The previous data from \cite{Lorenzen83} is presented for comparison. 
 The fitting procedure is explained in detail below and includes Rydberg transitions from \cite{Lorenzen83} for the 
transitions outside of the range of this experiment. Some re-analysis was required for the older data to take into account the changes in the recommended Rydberg constant $R_\textrm{Rb} = 10973660.672249(73)$ (using ref. \cite{codata06} and ref. \cite{Wapstra03}) since the publication of the previous paper.
From these results we obtain the fit results in table \ref{table:fit}. 
\begin{figure}[ht]
\begin{center}
\includegraphics[width=85.0mm]{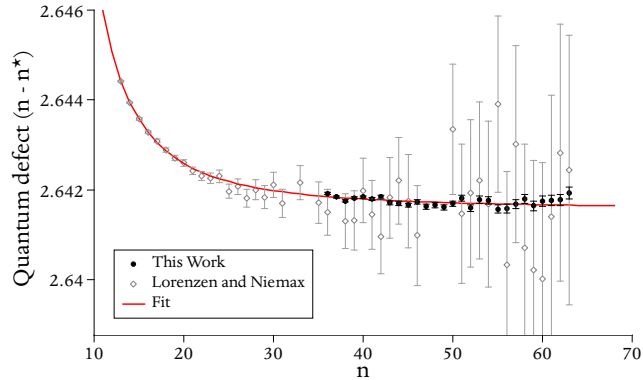}
\end{center}
\caption{Quantum defects}
\label{figure:defect_comparison}
\end{figure}

\section{Analysis}

Owing to subtleties in the fitting process that lead to subtle changes in the fitting parameters (see Table \ref{table:fit}), three methods are used to analyse the data. 

\noindent{\bf Method 1}: The first method uses a simple fit routine. 
The quantum defect can be obtained as a function of $n$ if we  approximate  $\delta(n)$ by $\delta_0$ \cite{Martin80}:
\begin{eqnarray}
	\delta(n) = \delta _0 + \frac{a}{(n-\delta _0)^2} + \frac{b}{(n-\delta _0)^4} + \cdots 
\label{eqn:ritz-extended}
\end{eqnarray}
in which case the equation can be fitted directly, results are shown in Table \ref{table:fit} (Method 1). 
The fitting constants obtained in this way lose their physical meaning \cite{Drake91} owing to differences in the fitting procedures. 
This means that the fitting parameters obtained in this manner cannot be directly compared across experiments.

However, these fitting parameters can be substituted into 
\begin{equation}
 E_n = Ei - R_\textrm{Rb} \left( n - \delta _0 - \frac{a}{(n-\delta _0)^2} -  \frac{b}{(n-\delta_0)^4} - \cdots \right)^{-2} 
\label{eqn:ritz-approx}
\end{equation}
to obtain approximate values of the energies across the manifold.

\noindent{\bf Method 2}: To align this analysis with that of Lorenzen and Niemax \cite{Lorenzen83} the second method is an iterative fit involving first equation \ref{eqn:ritz-extended} to obtain an accurate value of the ionization energy $E_i$ and then substituting this value as a fixed parameter in a subsequent fit of equation \ref{eqn:ritz-approx} .
 This method yields the results shown in Table \ref{table:fit} (Method 2) and Figure \ref{figure:defect_comparison}.

Previous experiments \cite{Lorenzen83,Meschede87} were not sufficient to resolve a discrepancy between these two fitting methods, but the increased accuracy of the current experiment exposed a one standard deviation offset, making the precise definition of the fit more important. 
Drake \cite{Drake91} discusses, in detail, the advantage of each method together with the physics that can be learned from more accurate measurements of the quantum defect, thus there is some merit in considering a method of extracting the fit parameters from the data directly.

\noindent{\bf Method 3}: The final and most accurate method is a fit using the absolute values of the measured transitions. A direct fit to the data is possible with a Modified Rydberg-Ritz formula obtained by expanding the effective quantum number $n^* = n-\delta(n)$ as follows:
\begin{eqnarray}
	n^* & = & n - \delta(n) \\
	   & = & n - \delta _0 - \frac{\delta _2}{(n-\delta(n))^2} - \frac{\delta _4}{(n-\delta(n))^4} - \cdots 
\label{eqn:ritz}
\end{eqnarray}
This creates an ``extended Rydberg-Ritz formula'' that can be rearranged to give the quantum defect $\delta(n)$ :
\begin{equation}
 \delta(n) = n - n^* = \delta _0 + \delta _2 t_n + \delta _4 t_n^2 +  \cdots 
 \label{eqn:ritz-defect}
\end{equation}
where
\begin{equation}
\nonumber t_n = \frac{1}{n - \delta(n)} = \frac{E_i - E_{n,j}}{R_{Rb}} 
\end{equation}
This is convenient as it allows the energy of the transitions to be extracted more accurately and extrapolated to all other principal quantum numbers $n$.
As the principle quantum number $n$ increases, the energy $E_n$ approaches the ionization energy $E_i$, making $t_n$ a very small number and the series quickly converges \cite{Sommerfeld20,Hartree28, Langer30}.
Moreover, it is possible to fit the equations using an iterative procedure and the result of this fit is shown in Table \ref{table:fit} (Method 3).
It is important to note that due to the small number of data-points within each Lorentzian, the usage of an accurate and numerically stable fitting routine is critical. Here we use the MINPACK set of algorithms \cite{more80} (a modified Levemberg-Marquardt algorithm \cite{Marquadt63}) initialized by accurate heuristics. 
The accuracy and stability of this fitting procedure was checked by fitting simulated data and compared very favorably against the nonlinear fitting and minimization algorithms present in commercial software packages. 
We put considerable effort in obtaining high quality experimental data (high signal to noise ratio and low dark count rate, as shown in Figure \ref{figure:typical_line}) which eliminates any ambiguity in the fitting.
Algorithms built around MINPACK were also used for fitting and minimization of the quantum defect in Equations \ref{eqn:ritz-extended} and \ref{eqn:ritz-defect}; although here the other algorithms also provided sufficient quality and stability.

\begin{table}[ht]
 \begin{center}
\begin{tabular}{c|lll}
 & Method 1 & Method 2 & Method 3 \\  \hline
$E_i$ & 1010024692(7) & 1010024692(7) & 1010024700(7) \\ 
$\delta_0$ & 2.641352 & 2.64149(1) & 2.64157(2) \\ 
a & 0.4822 & 0.317(3) & 0.304(4) \\ 
b & -2.86 & -7.45(5) & 1.15(5) \\ 
c & 11.3 & 1.9(2) & 1.2(2)
\end{tabular}
\end{center}
\caption{Result comparison using three different fitting methods: Method 1 uses Equation \ref{eqn:ritz-extended} exclusively. Method 2 uses $E_i$ found by fitting Equation \ref{eqn:ritz-extended} as a constant in Equation \ref{eqn:ritz-approx}. Method 3 uses Equation \ref{eqn:ritz-approx} exclusively. $E_i$ is quoted in MHz. The errors stated include systematic uncertainties. }\label{table:fit}
\end{table}



\section{Conclusion}

We have presented a precision measurement of the $n=36$ to $n=63$ P$_{3/2}$ Rydberg levels of ${}^{85}$Rb. We have analysed these results and found that we have made a significant improvement on the knowledge of the Modified Rydberg Ritz formula for ${}^{85}$Rb P-states. The results represent an order of magnitude improvement in the predictive capability for the $n$P$_{3/2}$ rubidium Rydberg state and we obtain an improved ionisation energy of $E_i$ =1~010~024~700(7)MHz compared with  $E_i$ = 1~010~024~684(90)MHz from reference \cite{Lorenzen83} and $E_i$ = 1~010~025GHz from reference \cite{Johansson61}.


\section{References}

\end{document}